\documentclass[aps,amsmath,amssymb,12pt]{revtex4}
\usepackage{graphicx}
\usepackage{bm}
\usepackage{tcolorbox}

\def\journal#1#2#3#4{{#1} {\bf #2}, #3 (#4)}
\newcommand{\be}{\begin{equation}}
\newcommand{\ee}{\end{equation}}
\newcommand{\bea}{\begin{eqnarray}}
\newcommand{\eea}{\end{eqnarray}}
\newcommand{\hf}{\frac12}
\newcommand{\nn}{\cr}
\def\eq#1{(\ref{#1})}
\def\la{\langle}
\def\ra{\rangle}
\def\tr{{\mathrm{tr}}}
\def\Tr{{\mathrm{Tr}}}
\def\ord#1{{\cal O}(#1)}
\def\mr#1{{\mathrm{#1}}}
\def\v#1{{\bm{#1}}}
\def\fd#1#2{\frac{\delta#1}{\delta#2}}
\def\fdd#1#2#3{\frac{\delta^2#1}{\delta#2\delta#3}}
\def\dk{{\Delta k}}
\def\hphi{\hat\phi}
\def\hchi{\hat\chi}

\def\hD{{\hat D}}

\def\sign{\mr{sign}}
 
\begin{document}
\title{Classical limit of a scalar quantum field theory}
\author{S. Nagy$^1$, J. Polonyi$^2$}
\affiliation{$^1$Department of Theoretical Physics, Faculty of Science and Technology, University of Debrecen, P.O. Box 400, H-4002 Debrecen, Hungary}
\affiliation{$^2$Strasbourg University, CNRS-IPHC, BP28 67037 Strasbourg Cedex 2, France}
\date{\today}
\begin{abstract}
It is well known that a minimal distance emerges in quantum field theories owing to the need to regularize the UV divergences. The macroscopical limit at large minimal distance, weak spatial resolution, is investigated for a self interacting scalar quantum field theory by the help of the renormalization group. The lowering of the cutoff always opens the dynamics hence the renormalization group has to be implemented for open quantum field theories. A strongly coupled non-relativistic scaling regime is found supporting a second order phase transition between weakly and strongly open theories. The weakly (strongly) open bare theories develop into strongly (weakly) open dynamics during the renormalization group flow. The two known conditions of classical limit, the strong decoherence and the suppression of the quantum fluctuations are confirmed for closed bare theories at distances beyond a non-relativistic correlation length.\end{abstract}
\maketitle

\section{Introduction}
Quantum and classical physics are based on seemingly contradictory concepts nevertheless the latter is supposed to follow from the former. The main difficulty of the derivation of classical physical laws stems from a qualitative difference, namely only the classical level possesses deterministic laws. The most natural way to address such a difference is look into the logical structure of the theories. It has actually been noticed long time ago that the Hilbert space based quantum logic is non-distributive as opposed to the set theory based Boolean logic of classical physics \cite{birkhoff,mackey}. However it is still an open question how such a difference may help us to solve the problem. In the absence of a formal solution one looks for more intuitive ideas. An interpolation between the quantum and the classical levels is established below rather than trying to match qualitatively different physics.

In a sharp contrast to the continuous structure of classical physics quantum systems possess elementary particles or excitations. The symmetries are represented in the linear space of quantum states as the linear sum of irreducible representations and the vectors of the latter stand for elementary particles. The seemingly continuous classical physics emerges where the observations can not resolve anymore the individual elementary particles. Thus the number of degrees of freedom unresolved by the observation, $N_{unr}$, can play the role of a parameter to locate the quantum-classical transition. The complication is that strongly correlated degrees of freedom are not really different hence we must find the number of independent degrees of freedom in this approach.

As soon as we retain the effects of the unresolved degrees of freedom we deal with an open system. Hence the quantum-classical transition can only be understood in terms of open systems. Two necessary conditions of the classical limit are known, the observed quantities must be decohered and display well determined values, without indeterministic quantum fluctuations. Naturally both conditions are to be realized only asymptotically, when $N_{unr}\to\infty$.

Decoherence stands for the suppression of the interference contributions in an expectation value between two components of a pure state \cite{zehd,zurekd,joos,zurekt}. The interference always refers to a contribution of two eigenstates of an observable in question. Hence the decoherence is an observable, or better to say a basis, dependent concept. The presence of interference contributions follows from the basic postulates of quantum mechanics and cannot be avoided in closed systems. However the question of interference becomes highly non-trivial when the Hilbert space of states is the direct product of the Hilbert spaces of an observed system and its unobserved environment because the interference of the observed factors is modulated by the unobserved environment. This is the origin of the environment induced decoherence: Let us consider a factorizable basis $|\psi_j\ra\otimes|\phi_j\ra$ for the full system where the first and the second component stand for the observed system and its environment, respectively. The overlap $\la\psi_j|\psi_k\ra$, relevant in a closed system, is actually $\la\psi_j|\psi_k\ra\la\phi_j|\phi_k\ra$ owing to the environment and decoherence arises when $|\la\phi_j|\phi_k\ra|<1$. Strong decoherence, expected in the classical limit, is supposed to arise from a soft environment where the relative environment states of different system states are approximately orthogonal.

Classical behavior is expected when the observation leaves a large number of elementary excitations, relevant for the observed quantity, unresolved. It appears therefore natural that the central limit theorem, recast in quantum mechanics \cite{macr}, reinforces the emergence of well defined physical quantities without fluctuations in the macroscopic limit, $N_{unr}\to\infty$. The unresolved degrees of freedom are supposed to be statistically independent when the central limit theorem is used, a condition which seems rather unphysical. But luckily a generalization of the central limit theorem for correlated probability variables is known, it is the renormalization group method. This method is used below to perform the limit $N_{unr}\to\infty$.

The basic idea of the renormalization group is to follow the change of the dynamics during the blocking, the successive elimination of the dynamical degrees of freedom \cite{kadanoff,wilson}. In other words, we modify the separation between the observed system and the environment by successively displacing degrees of freedom from the system to the environment. One encounters mixed states during this process hence we need an extension of the traditional blocking procedure, developed originally for classical statistical physics and generalized formally for the path integral of quantum transition amplitudes between pure states, in terms of the density matrix. The use of the Closed Time Path (CTP) scheme \cite{schw,keldysh} which covers both quantum, classical, closed and open dynamics \cite{cqco} appears perfectly suitable for this goal.

The treatment of many particles needs quantum field theoretical method hence we use here a simple $\phi^4$ scalar field theory model. There have been several works devoted to the renormalization group method within the CTP formalism. It was introduced by coarse graining in \cite{lombardo,dalvit,anastopoulos}, applied to quantum dots \cite{gezzi} and to open electronic systems \cite{mitra}. It can be used to describe transport processes \cite{jacobs}, damping \cite{zanellad}, critical dynamics \cite{bergerhoff,canet,mesterhazy,sieberer}, stochastic field theory \cite{zanellac}, and Bose-Einstein condensate \cite{gasenzerbe}. Applications in high energy physics extend over the one-loop renormalizability of the scalar model \cite{avinash}, inflation \cite{zanellai}, quantum cosmology \cite{calzettac} as well as over the use of spectral functions \cite{pawlowskisf}, the real time dynamics of gauge theories \cite{kasper}, nonthermal fixed points \cite{bergesfp} and the time dependence \cite{gasenzer}. The goal of the present work is to find the dependence of the decoherence and the fluctuations on the spatial resolution of a simple $\phi^4$ scalar model determined by the UV momentum cutoff by extending the results of \cite{rqft}.

Two aspects of cutoffs play a central role in this work. A gliding short distance cutoff is used to realize a flexible separation between the observed system and its environment. Furthermore, a cutoff for mass-shell singularities parametrizes the open interactions. Such unusual roles of the cutoffs motivate section \ref{divs} where the cutoffs of field theory models are revisited from our point of view. The main point is that the unavoidable short distance and mass-shell cutoffs make quantum field theories open. Open quantum field theories are briefly introduced in section \ref{openqft} by the help of the CTP formalism. A special attention is paid to the quadratic part of the action to avoid the restrictions of the asymptotic frequency condition, employed in closed theories. This procedure breaks the relativistic boost symmetry and section \ref{relboost} is included to argue that the system-environment entanglement makes any system-environment separation inherently non-relativistic in the quantum domain. The changing of the spatial resolution, the gliding UV cutoff separating the IR system and its UV environment, is introduced in section \ref{rgoqft}.

Section \ref{scalaw} contains the numerical results and the interpretation of the renormalization group flow: A second order phase transition is found between weakly and strongly open theories. What is remarkable is that this phase structure is inverted, namely theories with more closed bare parameters belong to the phase with more open long distance physics. Furthermore three crossovers were identified: The UV relativistic scaling laws go over non-relativistic scaling at a relativistic crossover. The open parameters increase sharply when a further, open interaction scale is reached within the strongly open phase. Finally a non-relativistic correlation length is found in the weakly open phase. The quantum fluctuations beyond this correlation length decrease according to the $\sqrt{N}$ law of the central limit theorem.

\section{Divergences and their removal}\label{divs}
It is widely known that quantum field theories are plagued by divergences. But it is less known that the regulators, introduced to handle such problems, generate open interactions. Since this feature is essential to our purpose a short outline of divergences and their elimination is presented in this section.

\subsection{Classification}
The divergences can be classified by considering the scale where they arise or the role they play. The best known ultraviolet (UV) divergences come from short distance or time. The simplest examples are provided by classical field theories dealing with point particles. In fact, a Feynman graph of $\ell$-loop is of $\ord{\hbar^{\ell-1}}$ where $\hbar$ is the Planck constant, indicating that there are classical one-loop graphs, owing to the $\ord{\hbar^{-1}}$ exponent of the path integral formalism. It is easy to find these graphs, they are generated when degrees of freedom are eliminated. For instance, the open interaction channels in the dynamics of charged particles belong to the electromagnetic radiation. The elimination of those degrees of freedom generates a one-loop (self energy) correction to the effective action for charged particles \cite{classcharge}. The UV divergences of field theories originate from the abundance of high energy modes in the Fock space. Such divergences are not restricted to field theories of point particles, there are short time divergences in 1+0 dimension, in quantum mechanics, where almost all trajectories of the path integral are nowhere differentiable fractals. The easiest way to see this is to perform a Wick rotation and to recall the similar property of a Brownian particle, described by the heat equation.

The long distance infrared (IR) divergences result from the piling up of some correlations, a typical example being the IR singular structure of the Coulomb field, to be softened by screening. The mass-shell (MS) divergences arise in the long observation time limit and indicate the softness of modes close to the mass shell, such as the divergence in the Green functions before the application of the $i\epsilon$ prescription. The MS divergences become more severe in the presence of massless particles and are called collinear \cite{bloch,kinoshita,lee}.

While we know the physics only in a finite length scale interval there is an obvious asymmetry in the way the unknown short and the long distance physics are treated. This comes from the general strategy to analyse complex systems: First one identifies their elementary constituents and after that one matches the elementary parts to reproduce the complexity of the full system. The key point is that the elementary constituents which appear with unknown properties tend to be smaller than the whole system. Hence once the UV physics is clarified the IR prediction of the theory are supposed to be physical. Therefore a divergence can be physical or unphysical and physical theories are supposed to handle only the former. The UV divergences are unphysical, they arise from extrapolating physical laws to short distances where they have not yet been confirmed experimentally. The MS and IR divergences correspond to physical phenomenas, examples for the former are the time arrow and the soft photon collinear divergences, the latter can signal an instability leading to a long range rearrangement of the vacuum, a phase transition in the thermodynamical limit.

\subsection{Regularization}
The models are useful only if they are free of divergences thus one has to introduce some regulator, called cutoff, to render the mathematical structure well defined. The cutoffs are artificial formal devices and must be removed before confronting the prediction of the theory with observations.

A UV cutoff, the momentum cutoff expressed by the inequality $|\v{p}|<\Lambda$ in this work, is called sharp when it restricts the Fock space and represents a  definition of the degrees of freedom. A smooth cutoff, such as the Pauli-Villars regulator, leaves the Fock space unchanged but modifies the dynamics, typically the free dispersion relation, to make the dynamics UV  finite. The smooth analytical regularizations rely on the analytical extension in some parameter, e.g. the dimension of the space-time, in such a manner that the divergences are recovered only when the physical value is approached. The UV cutoff-dependence represents our ignorance of physics and is swept under the rug by making the theory independent of its numerical value. This can be made by the blocking procedure \cite{kadanoff,wilson}, where a decrease of $\Lambda$ is followed by a redefinition of the parameters of the dynamics in such a manner that the physics is kept $\Lambda$-independent. A fixed UV sharp cutoff $\Lambda$ restricts the Fock space and thereby introduces observable, physical effects. The gliding cutoff used in the blocking procedure does not appear in the dynamics and will be denoted by $k$ in this paper.

An IR cutoff can be spatial or temporal. The spatial IR divergences, if present, are usually regulated by the introduction of a finite quantization box in space. Physically tested non-gravitational theories should possess convergent thermodynamical limit when all important long range correlation are accounted. Finally, the MS singularities can simply be avoided by a temporal IR cutoff which renders the observed frequencies sufficiently vague.

The cutoffs play a surprisingly complicated role if their removal generates non-uniform convergent integrals \cite{spont}. The non-uniform convergent loop-integrals during the removal of the UV cutoff give rise to anomalies in the perturbations series. The non-uniform convergence in the thermodynamical limit reveals the presence of slow long distance modes which leads to non-commuting thermodynamical and long time limits and indicates phase transitions. Finally, the non-uniform convergence of the Fourier integral when the poles are slightly shifted from the real integration contour in the complex frequency plane leads to breakdown of the time reversal symmetry and introduces a causal time arrow.

\subsection{MS divergences}
The properties and the role of the MS singularities are not widely discussed. The formal origin of these singularities is an improper handling of a discrete spectrum value embedded in the continuum. To gain more physical insight let us consider a degree of freedom $x$ equiped with a dynamics, defined by the action
\be
S_0[x]=\int dt\left[x(t)K(\partial_t)x(t)+j(t)x(t)\right]
\ee
where $K(z)$ is a polynomial of finite order with real coefficients. The null-space of the operator $K(\partial_t)$ consists of the solutions of the homogeneous equation of motion. These modes have ill-defined dynamics when the inhomogeneous equation of motion, $K(\partial_t)x(t)=j(t)$ is solved. The usual solution is to exclude these modes from the dynamics and to suppress the source in the null-space. But these modes have to be fixed, as well, that is achieved ultimately by the auxiliary conditions supplied besides the equation of motion. Hence we need a special treatment of the null-space, few discrete modes embedded into the continuous spectrum of the Fourier representation of the trajectory $x(t)$. There are different null-spaces for each wave vector in field theory and those make up the mass-shell modes. The MS singularities are bound up with IR divergences because the mass-shell has continuous spectrum in the thermodynamical limit.

The natural way to avoid MS singularities is to restrict the allowed space-time region to a finite size in such a manner that the resulting discrete frequency spectrum skips the null-space eigenfrequencies. One can not follow relaxation process in a finite space-time regions therefore one prefers instead to keep the space-time unrestricted and to employ an analytical regularization, the $i\epsilon$ prescription: An infinitesimal imaginary term is introduced in the free action, $K\to K\pm i\epsilon$, where $\epsilon$ plays the role of a cutoff, parametrizing the way the discrete spectrum contribution is spread over the continuum. The MS regulator solves another problem, as well, it renders the path integral over the trajectories convergent without Wick rotation.

\subsection{Cutoffs and environment}
The cutoffs play yet another key role, the UV and the MS regulators imply and represent environment, respectively. An UV cutoff was originally introduced to remove the short distance modes with unknown dynamics. But these modes are physical and are suppressed in a measurement only within the observing apparatus. In other words, the degrees of freedom residing beyond the UV cutoff make up an unobserved environment and render the observed dynamics open. Note that only sharp cutoff can be used for such a purpose.

One customarily places the UV cutoff far away from the observed scales to make its detail unimportant and relies on the approximation where we simply ignore the impact of the UV environment, cf. sharp momentum cutoff. We can use closed field theory models in this case with the price of treating the cutoff as a physical parameter. As soon as we try to hide it by the blocking the dynamics of the observed modes become open \cite{rqft}. Though the non-physical gliding momentum space cutoff $k$ does not influence the physical content the spatial resolution it determines the minimal distance of the theory, $\ell_m=1/k$. The UV environment modes which make the observed dynamics open are actually the origin of the scale dependence of physical laws, c.f. section \ref{opnincl}.

One may object that the environment of the high energy modes are not important at low energy hence the openness of the blocked dynamics plays not much role neither. In other words, the UV environment is hard for large value of the cutoff and is difficult to excite. In fact, the contributions of high energy modes in the Rayleigh-Schrödinger stationary perturbation expansion tend to be suppressed by the large denominator. But such a suppression holds only in the first quantized quantum mechanical models which describe the physics in a relatively narrow scale regime. The quantum field theories are supposed to cover a much wider scale windows and the UV divergences of the perturbation expansion indicate that the large number of high energy intermediate states overwrites the suppression of the large denominators.

The proper handling of the troublesome dominating UV contributions of the unobserved scales is provided by the renormalization group method. The resulting picture of universality can be summarized by the statement that the renormalizable parameters of a theory provide a natural parametrization of the manifold of possible low energy physics as long as there is a sufficiently long UV scaling regime, cf. \cite{grg} for some important exceptions. It turns out that there are renormalizable open interaction vertices and they parametrize the {\em cutoff-independent and unavoidable} open interactions of low energy physics.

The MS regulator represents an environment but in a different manner than a sharp UV cutoff. The $i\epsilon$ prescription regulating the MS singularities actually stands for an infinitesimal time reversal symmetry breaking open interaction, c.f. the remark after eq. \eq{sbc}. This can be understood intuitively within a time reversal invariant closed dynamics by raising the question about the origin of the causal time arrow. The latter points in the direction of time in which the effects of a time-localized external perturbation are observed and a time reversal invariant dynamics possesses no causal time arrow by definition. Nevertheless, according to our daily experience such a time arrow exists in our world where the weak CP breaking of the weak interactions can be neglected. The time arrow is introduced in our equations by the choice of the auxiliary conditions for the equations of motion. In fact, by specifying the auxiliary conditions to our differential equations locally in time one generates a causal time arrow which points {\em away} from the time of the auxiliary conditions. There are several causal time arrows in physics \cite{zeh,halliwell}, all induced by some environment. The $i\epsilon$ prescription pushes off the poles of the propagator from the integration contour on the complex frequency place and the sign of $\epsilon$ stands for a causal time arrow as soon as the frequency integrals are calculated by means of the residuum theorem. This time arrow must have come from some physical environment, not retained in our calculation.

The relation of the regularization of the MS singularities and open dynamics is particularly important point because it shows the impossibility to define a closed theory for arbitrarily long observation time. The initial condition for the integration of the beta functions in the renormalization group treatment, the bare theory at a large but finite UV cutoff, must have open interaction channels with finite strength.

An obvious but important result of the unavoidable presence of the MS regulator is the loss of the Gaussian fixed point because the theory is ill defined for vanishing $\epsilon$. One could argue that this is not very important because the limit $\epsilon\to0$ is convergent and is anyhow  carried out at the end of the calculation. But the imaginary term in the action breaks time reversal invariance in an infinitesimal manner and its impact remains finite, in a manner reminiscent of spontaneous symmetry breaking \cite{spont}. Such an interpretation is trivial for a harmonic oscillator but becomes highly involved in the presence of interaction, cf. the existence of a separatrix of the renormalization group flow mentioned in section \ref{openptrs}.

\subsection{Open interactions in a closed cutoff theory}\label{opnincl}
The discussion of open interaction channels in a quantum field theory model is unnecessary when the model is  defined by the help of fixed, physical, sharp UV cutoff. The point here is that the same phenomenon can be described in terms of either closed or open interactions depending of the choice of degrees of freedom used in our equations.

Perhaps the simplest way to demonstrate this point is to compare exclusive and inclusive scattering processes. All particles participating in the scattering process are supposed to be detected in the former and only some of the final particles are identified in the latter. The undetected particles represent an environment in the theoretical description of an inclusive scattering process. The formal framework of the difference between exclusive and inclusive processes can be found by considering the harmonic oscillators representing the free particles with a fixed momentum in canonical quantization. The interaction is closed when expressed in terms of states of the Fock space which specifies the state of all oscillator. But as soon as some oscillators are treated unobservable the dynamics for the rest is open. Translated into particle scattering language, one is dealing with closed dynamics only in exclusive processes which are meaningful if there is a sufficiently large mass gap to exclude undetected particles in a measurement with small but finite energy-momentum uncertainties.

The application of the optical theorem to the Schwinger-Dyson partial resummation of the perturbation series for a propagator can serve as a simple example of the emergence of open interaction channels in a closed cutoff theory. One separates the forward scattering in the scattering matrix by the introduction of the $T$ matrix, $S=\openone+iT$, and the optical theorem expresses the unitarity of the time evolution in a closed dynamics, $S^\dagger S=\openone$, by the equation
\be\label{opt}
i(T^\dagger-T)=T^\dagger T.
\ee
An important implication of this relation is that Feynman graphs with on-shell external legs assume complex value. It follows that the self energy acquires finite imaginary part on the mass-shell indicating that that a sufficiently high energy particle may be unstable and part of the state norm decays into other particle channels which appear as an environment of the one-particle dynamics. Unitarity is lost within the one-particle sector but the conservation of the the total probability can be established by representing the state of the particle by a density matrix. The environment encoded by the Schwinger-Dyson partial resummation is reminiscent of polarons in solids and this analogy corroborates the view of the dressing of a bare particle, its polarization cloud, as the result of open interactions.

Another example is provided by the BKLN theorem \cite{bloch,kinoshita,lee} stating that one can observe only inclusive processes in a gauge theory without gap. The key idea, the cancellation between the on-shell real and the off-shell virtual photons, is actually more natural and even easier to follow in the CTP formalism for closed QED where it corresponds to the cancellation between the leading IR terms of the closed and the open interactions from the point of view of the harmonic oscillator of the given momentum sector in the framework of the Schwinger-Dyson partial resummation.

\section{Open quantum field theories}\label{openqft}
The basic steps of the construction of open quantum field theories are outlined in this section.

\subsection{Density matrix in a closed dynamics}
Let us start with a real scalar field $\phi(x)$ and the path integral expression for the transition amplitude between pure field eigenstates
\be\label{closed}
\la\phi_f(\v{x})|e^{-i(t_f-t_i)H}|\phi_i(\v{x})\ra=\int_{\phi(t_i,\v{x})=\phi_i(\v{x})}^{\phi(t_f,\v{x})=\phi_f(\v{x})}D[\phi]e^{iS[\phi]},
\ee
where the integration extends over trajectories with given initial and final field configuration. The units $c=\hbar=1$ are used below. The traditional formalism with such a path integral expression is called Single Time Path (STP) formalism below. To recover the time translation invariance one performs the limit $t_i\to-\infty$, $t_f\to\infty$ and employs the usual $i\epsilon$ prescription
\be
S[\phi]\to S[\phi]+i\frac\epsilon2\int dx\phi^2(x).
\ee
The UV regulator, a sharp cutoff in momentum space, $|\v{p}|<\Lambda$, is included into the definition of the integration measure. It is important to keep in mind that the quantum bare action occurring in the path integral is the classical action, cf. section \ref{subtrs}. We shall see that the quantum action for open theories is complex in contrast to the real classical action.

In preparing the way for open dynamics we seek the density matrix in path integral form,
\bea
\rho[\phi_+(\v{x}),\phi_-(\v{x})]&=&\la\phi_+(\v{x})|e^{-i(t_f-t_i)H}\rho_ie^{i(t_f-t_i)H}|\phi_-(\v{x})\ra\nn
&=&\int_{\phi_\pm(t_f,\v{x})=\phi_\pm(\v{x})}D[\hphi]e^{iS[\hphi]}\rho_i[\phi_+(t_i,\v{x}),\phi_-(t_i,\v{x})],
\eea
where the field doublet $\hphi=(\phi_+,\phi_-)$ has been introduced for the trajectories contributing to the matrix elements of the time evolution operator for the bra and the ket components together with the action $S[\hphi]=S[\phi_+]-S^*[\phi_-]$.

The density matrix of a pure state is factorizable, $\la\phi_+|\rho|\phi_-\ra=\rho[\phi_+,\phi_-]=\Psi[\phi_+]\Psi^*[\phi_-]$, meaning that that the quantum fluctuations in the bra and the ket components are factorizable. Such a factorization is preserved during the time evolution due to the additive structure of the action $S[\hphi]$.

\subsection{Reduced density matrix for a subsystem within a closed dynamics}
Let us denote the field variables of the observed system and its environment by $\phi$ and $\chi$, respectively and assume that the closed full dynamics is described by the bare action $S[\phi,\chi]$. The transition amplitude between pure field eigenstates is of the form
\bea
{\cal A}&=&\la\phi_f(\v{x}),\chi_f(\v{x})|e^{-i(t_f-t_i)H}|\phi_i(\v{x}),\chi_i(\v{x})\ra\nn
&=&\int_{\phi(t_i,\v{x})=\phi_i(\v{x})}^{\phi(t_f,\v{x})=\phi_f(\v{x})}D[\phi]\int_{\chi(t_i,\v{x})=\chi_i(\v{x})}^{\chi(t_f,\v{x})=\chi_f(\v{x})}D[\chi]e^{iS[\phi,\chi]},
\eea
where the integration extends over trajectories with given initial and final spatial field configuration. The reduced density system matrix
\be
\rho[\phi_+(\v{x}),\phi_-(\v{x})]=\la\phi_+(\v{x})|\Tr_\chi\left[e^{-i(t_f-t_i)H}\rho_ie^{i(t_f-t_i)H}\right]|\phi_-(\v{x})\ra
\ee
where the trace is over the environment Fock space. The initial density matrix is assumed to be factorizable, $\rho_i=\rho_{i\phi}\rho_{i\chi}$, c.f. section \ref{relboost}.

The reduced system density matrix can be written in the form of a path integral,
\be
\rho[\phi_+(\v{x}),\phi_-(\v{x})]=\int_{\phi_\pm(t_f,\v{x})=\phi_\pm(\v{x})}D[\hphi]e^{iS[\hphi]}\rho_i[\phi_+(t_i,\v{x}),\phi_-(t_i,\v{x})],
\ee
where the action is defined by the equation
\be\label{elim}
e^{iS[\hphi]}=\int_{\chi_+(t_f,\v{x})=\chi_-(t_f,\v{x})}D[\hchi]e^{iS[\hphi,\hchi]}\rho_e[\chi_+(t_i,\v{x}),\chi_-(t_i,\v{x})]
\ee
with $S[\hphi,\hchi]=S[\phi_+,\chi_+]-S^*[\phi_-,\chi_-]$. The Hermiticity of the density matrix implies the equation
\be\label{herm}
S[\phi_+,\phi_-]=-S^*[\phi_-,\phi_+].
\ee

To underline the new features of the expressions leading to the density matrix rather than the transition amplitude one may consider the total probability,
\be
\Tr[\rho]=\int_{\phi_+(t_f,\v{x})=\phi_-(t_f,\v{x})}D[\hphi]e^{iS[\hphi]}.
\ee
where the convolution with the initial density matrix is suppressed, a convention to be employed below. This scheme is called Closed Time Path formalism due to the integration over the common final environment configuration representing the trace over the system Fock space. The distinguished feature of this formalism is the formal reduplication of the degrees of freedom, $\phi\to\hphi$, is actually a remarkable efficient way to represent the environment \cite{cqco}.

The coupling of the trajectories of the two environment copies at the final time in \eq{elim} couples the two system copies, as well, thereby introducing system-environment entanglement and rendering the system density matrix non-factorizable. The imaginary part of the $S[\hphi]$ controls decoherence in the field eigenstate basis and the convergence of the path integral requires $\mr{Im}~S[\hphi]>0$.

The expectation value of the time ordered field-dependent observable $O[\phi]$ can be obtained by inserting the functional $O[\phi]$ with either field variable,
\be\label{pmua}
\la T[O[\phi]]\ra=\int D[\hphi]O[\phi_\pm]e^{iS[\hphi]}
\ee
and the equation remains independent of $t_f$ which is supposed to be later than any time occuring in the observable.

\subsection{Conditions on the action}\label{conds}
The action $S[\phi_+,\phi_-]$ characterizing the open dynamics is rather involved since far more elementary processes, vertices, may be present for the CTP field doublet than for the single field variable of the traditional STP formalism. It is therefore more important to find the conditions restricting the phenomenologically motivated action functionals.

This is a well known and simple issue for closed theories where the necessary condition is that the Schrödinger equation corresponding to the path integral should lead to a Hermitian Hamiltonian. The derivation of the equation of motion from the path  integral is rather straightforward: (i) Define the wave functional $\Psi[t_f,\phi_f(\v{x})]$ as given by  eq. \eq{closed}. (ii) The path integral is based on discretized time, $t_n=n\Delta t$. Consider the last infinitesimal time step $t_f-\Delta t\to t_f$ and the corresponding part of the path integral as a linear map of the wave functional $\Psi[t_f-\Delta t,\phi'(\v{x})]\to\Psi[t_f,\phi(\v{x})]$ and expand the integrand around the free, harmonic theory using $\Delta t$ as small parameter. Such a single time step evolution restricts the jump of the field configuration to $\ord{\sqrt{\Delta t}}$, allowing to use a Taylor expansion of $\Psi[t-\Delta t,\phi(\v{x})]$ in $\phi'(\v{x})-\phi(\v{x})$. The result is that real Lagrangians which are additive in a quadratic kinetic and the potential energy produce Hermitian Hamiltonian.

Similar steps can be followed to derive the master equation for the density matrix \cite{timescale} but this latter must fulfill more involved conditions, unit trace, Hermiticity and positivity. While the verification of the preservation of the trace and the Hermiticity is rather trivial the positivity is difficult to assure. A possible solution is to reconstruct some kind of environment which could have resulted the master equation in question. When this is possible then the master equation is called of Lindblad form  \cite{lindblad}. It is impossible to guess the Hilbert space of an unknown environment hence one follows an alternative route, based on the linear space ${\cal L}$ of operators span by the open interaction terms in the master equation. The master equation preserves the positivity if it can be written as the sum of the commutator with a Hamiltonian, the closed Neumann equation, and a dissipator which is a particular quadratic expression of the open interaction operators with a Hermitian and positive matrix as coefficient. Since ${\cal L}$ is low dimensional it is usually not difficult to assure the positivity of the coefficient matrix. The verification of the Lindblad structure of the scalar theory has been made only in the leading order of the perturbation expansion \cite{avinash}.

We impose the simpler condition \eq{pmua} restricted to the field operator. This condition is expressed in terms of the effective action,
\be
\Gamma[\hphi]=W[\hat j]-\int dx\hat j(x)\hphi(x),
\ee
obtained from the generator functional for connected Green functions,
\be
e^{iW[\hat j]}=\int D[\hphi]e^{iS[\hphi]+i\int dx\hat j(x)\hphi(x)}
\ee
by a functional Legendre transformation where $\hphi(x)=\delta W[\hat j]/\delta\hat j(x)$ is the field expectation value in the presence of the source $\hat j$. The field expectation value solves the variational equation of motion
\be\label{eeom}
\fd{\Gamma[\hphi]}{\hphi(x)}=-\hat j(x).
\ee
We impose the condition $\phi_+(x)=\phi_-(x)$ for the physical case $\hat j=0$. It is easy to check that the effective action satisfies the Hermiticity condition \eq{herm} and therefore $\phi_+(x)=\phi_-^*(x)$ and we actually require only that the field expectation be real.

Our condition is necessary for the positivity which can be seen in the leading order of expansion in the Planck constant. The positivity of the density matrix, the validity of the inequality $\la\Psi|\rho|\Psi\ra\ge0$ for any $|\Psi\ra$, implies the bound
\be
0\le\la\phi_f|\rho|\phi_f\ra=\int D[\hphi]\prod_{\v{x}}\delta(\phi_+(t_f,\v{x})-\phi_f(\v{x}))\delta(\phi_-(t_f,\v{x})-\phi_f(\v{x}))e^{iS[\hphi]}
\ee
for any $\phi_f$ and initial density matrix which is diagonal in the field eigenstates. The evaluation of the inequality on the tree level of the path integral yields
\be
0\le Ne^{iS[\hphi]}
\ee
where $N>0$ is a normalization constant. According to \eq{herm} $\mr{Re}~S[\hphi]=0$ when $\phi_+(x)=\phi_-(x)$ and the inequality is fulfilled.

\subsection{Free propagator}
It is advantageous to interrupt the discussion of open dynamics by the introduction of the propagators defined by the generator functional
\bea\label{genfree}
e^{iW[\hat j]}&=&\Tr T[e^{-i\int_{t_i}^{t_f}dt'[H(t')-j_+(t')\phi(t')]}]|0\ra\la0|
\bar T[e^{i\int_{t_i}^{t_f}dt'[H(t')+j_-(t')\phi(t')]}]]\nn
&=&\int D[\hphi]e^{\frac{i}{2\hbar}\int dxdx'\hphi(x)\hD^{-1}(x-x')\hphi(x')+i\int dx\hat j(x)\hphi(x)}\nn
&=&e^{-\frac{i}{2\hbar}\int dxdx'\hat j(x)\hD(x-x')\hat j(x')}
\eea
with independent external source for the bra and the ket components, $\hat j=(j_+,j_-)$, as
\be
D_{\sigma,\sigma'}(x,x')=\fdd{W[\hat j]}{ij_\sigma(x)}{ij_{\sigma'}(x')}.
\ee
The calculation of the propagator is straightforward for a free relativistic field with the action
\be
S_0=\hf\int dx[\partial_\mu\phi(x)\partial^\mu\phi(x)-m^2\phi^2(x)],
\ee
in the long time limit $t_i\to-\infty$, $t_f\to\infty$, where one recovers the Feynman propagator and the Wightman functions in the diagonal and the off-diagonal CTP blocks of the $2\times2$ block matrix
\be
i\hD(x,x')=\begin{pmatrix}\la T[\phi(x)\phi(x')]\ra&\la\phi(x')\phi(x)\ra\cr\la\phi(x)\phi(x')\ra&\la T[\phi(x')\phi(x)]\ra^*\end{pmatrix}.
\ee
The Fourier transform of translation invariant operators,
\be
f(x-y)=\int\frac{dp}{(2\pi)^4}e^{-ip(x-y)}f(p)
\ee
gives $\hD(p,p')=(2\pi)^4\delta(p-p')\hD(p)$ with
\be\label{freepro}
\hD(p)=\begin{pmatrix}\frac1{p^2-m^2+i\epsilon}&-2\pi i\delta(p^2-m^2)\Theta(-p^0)\cr-2\pi i\delta(p^2-m^2)\Theta(p^0)&-\frac1{p^2-m^2-i\epsilon}\end{pmatrix}.
\ee

It is advantageous to parametrize the propagator by three real space-time functions,
\be\label{propblock}
\hD=\begin{pmatrix}D^n+iD^i&-D^f+iD^i\cr D^f+iD^i&-D^n+iD^i\end{pmatrix}
\ee
where $D^i=-\pi\delta(p^2-m^2)$, $D^n(p)=P1/(p^2-m^2)$, $P$ standing for the principal value, and $D^f=-i\pi\delta(p^2-m^2)$. The physical significance of $D^n$ and $D^f$ can be discovered by the introducing a physical external source, $j_\pm=\pm j$, $S_0\to S_0+\int dx\hat j\hphi$ and inspecting the equation of motion \be
\phi_\pm(x)=\int dy[D_{\pm+}(x-y)-D_{\pm-}(x-y)]j(y).
\ee
One finds the causal propagators $D^{\stackrel{r}{a}}=D^n\pm D^f$ allowing us to identify $D^n$ and $D^f$ with the near and the far field components, respectively. We have the same real part of the CTP propagator in classical and quantum physics owing to Ehrenfest theorem which is exact for harmonic system  and the imaginary part $D^i$ enters only in the quantum case where it yields the spectral function $\rho(p)=iD_{-+}(p)=2\pi\delta(p^2-m^2)\Theta(p^0)$.

\subsection{Long time limit}
To recover time translation invariance one performs the limit $t_i\to-\infty$, $t_f\to\infty$ in \eq{closed}. The former limit poses no problem since the vacuum state is time translation invariant. However the latter is highly nontrivial due to the closing of the time path at the final time. As a preparation to perform the limits we introduce first a regulator for the MS divergences in \eq{elim} for finite $t_i$ and $t_f$,
\be
S[\phi,\chi]\to S[\phi,\chi]+i\frac\epsilon2\int dx[\phi^2(x)+\chi^2(x)].
\ee
Besides assuring the finiteness of the path integral and removing the null-space of the quadratic part of the action this step allows the elimination of the environment degrees of freedom by introducing an environment time arrow. The limit $t_f\to\infty$ can now be carried out for a harmonic system already on the classical level \cite{spont}.

We follow here a shorter and more conventional way in the quantum case, by defining first the path integral \eq{genfree} for a free, harmonic action with a kernel given by the inverse propagator. The replacement of the Dirac-delta with a Lorentzian,
\be
\delta_\epsilon(x)=\frac\epsilon{\pi(x^2+\epsilon^2)},
\ee
yields the inverse propagator
\be\label{invfreepr}
\hD^{-1}(p)=\begin{pmatrix}p^2-m^2+i\epsilon&-2i\epsilon\Theta(-p^0)\cr-2i\epsilon\Theta(p^0)&-p^2+m^2+i\epsilon \end{pmatrix}.
\ee
Thus the free action with the generalized $i\epsilon$ prescription is defined as $S[\hphi]\to S[\hphi]+S_{BC}[\hphi]$ with
\be\label{sbc}
S_{BC}=i\epsilon\int dp\left[-\sign(p^0)\phi_-(-p)\phi_+(p)+\hf|\phi_+(p)-\phi_-(p)|^2\right].
\ee
Note that the MS regulator which renders the Green function finite introduces an infinitesimal coupling to an environment since the first and the second terms of the right hand side of eq. \eq{sbc} represent an open force and decoherence, respectively. It is instructive to see the space-time expression
\be
S_{BC}=\epsilon\left[\frac1\pi\int d^3xdtdt'\frac{\phi_-(t,\v{x})\phi_+(t',\v{x})}{t-t'+i\epsilon}+\frac{i}2\int dx[\phi_+^2(x)+\phi_-^2(x)]\right]
\ee
which shows that the generalized $i\epsilon$ prescription goes beyond the $i\epsilon$ regulator of the  Feynman propagator by the introduction of an infinitesimal non-local in time open interaction. Such a quadratic action allows us to define the bare action of the path integral for anharmonic dynamics as $S[\hphi]\to S[\phi_+]-S[\phi_-]+S_{BC}[\hphi]$ in agreement with the restriction \eq{herm}.

\subsection{Asymptotic frequency conditions and causality}
The generalized $i\epsilon$ prescription includes the asymptotic frequency conditions, namely that all excitations above the vacuum are with positive or negative frequency for asymptotically early or late time, respectively. These conditions follow from the stability of the vacuum in closed dynamics and allow the adiabatic switching on the interaction, needed for the application of perturbation expansion for the vacuum-to-vacuum transition amplitude in the STP formalism.

However the asymptotic frequency conditions are violated in open dynamics. This is obvious in the canonical ensemble where the energy of a thermally distributed particle system can be either increased or decreased. In general, the system-environment interaction generate system-environment entanglement and the presence of several factorizable system-environment pure state components in the initial density matrix make it possible to increase or to decrease the energy asymptotically. Thus $\epsilon$ should be treated as a small but finite parameter.

The $i\epsilon$ prescription refers to the bare, regulated dynamics whose parameters characterize the processes around the minimal distance, the UV cutoff scale. Since the $i\epsilon$ prescription involves quadratic, super-renormalizable operators their renormalization is of central importance in the physics of finite scales. A major obstacle in treating $\epsilon$ small but finite is that the use of the MS regulator \eq{sbc} with finite $\epsilon$ generates acausalities owing to the non-analytical energy-dependence. Thus we need an alternative MS regulator which can reproduce \eq{sbc} for infinitesimal $\epsilon$ but remains causal for finite strength. Such a modification of the generalized $i\epsilon$ prescription is introduced in the next section, it allowed us to go beyond the results reported in the earlier work \cite{rqft}.

\subsection{A phenomenological action}
An alternative MS regulator is offered by the derivative expansion, the expansion of the action in momentum space. The most general quadratic form of the action with Lindblad structure \cite{lindblad} is of the form
\be\label{diblock}
\hD^{-1}=\begin{pmatrix}K^n+iK^i&K^f-iK^i\cr-K^f-iK^i&-K^n+iK^i\end{pmatrix}
\ee
with
\bea\label{fintuneda}
K^n(\omega,\v{p})&=&z\omega^2-\omega_p^2,\nn
K^f(\omega,\v{p})&=&i\nu\omega-i(d_0+d_2\omega^2),\nn
K^i(\omega,\v{p})&=&d_0+d_2\omega^2,
\eea
up to $\ord{\omega^2}$ and total time derivatives \cite{oho}. Here $\omega_p=\sqrt{m^2+\v{p}^2}$ and $m^2$, $\nu$, $d_0$, as well as $d_2$ are positive functions of $\v{p}^2$ satisfying $\nu^2\le4d_0d_2$. We employ below two different simplifications of the momentum-dependence. The simplest is to suppress it completely and use momentum-independent $z$, $\nu$, $d_0$ and $d_2$. A slightly more physical choice is where only $z$ and $d_0$ are kept momentum-independent and the remaining two parameters acquire a prescribed momentum-dependence, $\nu=\tilde\nu/\omega_p$, $d_2=\tilde d_2/\omega_p^2$. This choice is justified by noting that $\tilde\nu=d_0=\tilde d_2=g^2$ reproduces the closed free propagator \eq{freepro} as $g\to0$. The momentum-dependent parameters are used in the calculations leading to the Figures below.

To see better the physical content we record the Lagrangian with momentum-independent $z$, $\nu$, $d_0$ and $d_2$,
\bea\label{phena}
L&=&\hf[z(\partial_0\phi_+)^2-(\v{\nabla}\phi_+)^2-z(\partial_0\phi_-)^2+(\v{\nabla}\phi_-)^2-m^2(\phi_+^2-\phi_-^2)\nn
&&+\nu(\phi_-\partial_0\phi_+-\partial_0\phi_+\phi_-)+id_0(\phi_+-\phi_-)^2+id_2(\partial_0\phi_+-\partial_0\phi_-)^2],
\eea
The equation of motion reveals that $\nu$ represents a friction force in the internal space. The parameters $d_0$ and $d_2$ in the imaginary part parametrize the decoherence in the field diagonal basis. The inverse of \eq{diblock} is of the form \eq{propblock} and elementary calculation results the time dependence
\bea\label{freeprop}
D^i(t,\v{p})&=&-\frac{e^{-\frac\nu{2z}t}}{2z\omega_p^2}\begin{cases}\frac{zd_0+d_2\omega_p^2}\nu\cos\frac{\omega_p}zt+\frac{zd_0-d_2\omega_p^2}{2\omega_p}\sin\frac{\omega_p}zt&z\omega_p^2>\frac{\nu^2}4\cr
\frac{zd_0+d_2\omega_p^2}\nu\cosh\frac{\omega_p}zt+\frac{zd_0-d_2\omega_p^2}{2\omega_p}\sinh\frac{\omega_p}zt&z\omega_p^2<\frac{\nu^2}4\end{cases}\nn
D^{\stackrel{r}{a}}(t,\v{p})&=&-\frac{e^{-\frac\nu{2z}t}}{\omega_p}\begin{cases}\sin\frac{\omega_p}zt&z\omega_p^2>\frac{\nu^2}4\cr
\sinh\frac{\omega_p}zt&z\omega_p^2<\frac{\nu^2}4\end{cases}
\eea
with $\omega_p=\sqrt{|z\omega_0^2-\nu^2/4|}$ for underdamped $z\omega_p^2>\nu^2/4$ (overdamped $z\omega_p^2<\nu^2/4$) harmonic oscillator corresponding to the momentum $\v{p}$. The causal propagator agrees with the classical case. An important lesson is that the dissipation time scale of the free quasi-particles is given by the friction constant $\nu$, independently of decoherence. The spectral function is
\be
\rho(\omega)=\sign(\omega)[1+2n(\omega)]\pi\delta_{|\omega|\nu}(z\omega^2-\omega_p^2),
\ee
with
\be\label{npn}
n(\omega)=\frac{zd_0+d_2\omega^2}{2\omega\nu}
\ee
and the relative violation of the asymptotic frequency condition,
\be
\frac{D_{-+}(-\omega_p,\v{p})}{D_{-+}(\omega_p,\v{p})}=\frac{2n_0-1}{2n_0+1}
\ee
where $n_0=n(\omega_p)$

Assuming that the full closed dynamics of the system and its environment is known the open parameters can be calculated order-by-order in the perturbation expansion in the system-environment interaction strength $g$ and $z-1$, $\nu$, $d_0$ and $d_2$ arise in the leading, $\ord{g^2}$ order \cite{diss}. It is instructive to consider the weak coupling limit, $g\to0$, revealing the universal weak environment coupling limit
\bea
D^{\stackrel{r}{a}}(t)&\to&-\Theta(t)\frac{\sin t\omega_p}{\omega_p}\nn
D^i(t)&\to&-\frac{n_0}{\omega_p}e^{-i\omega_p|t|}.
\eea
While the weak coupling limit removes dissipation from the causal propagator the imaginary part of the propagator retains a finite, $\ord{g^2}/\ord{g^2}=\ord{g^0}$ contribution \eq{npn}. The peak of the spectral function becomes sharp but its weight and the violation of the asymptotic frequency condition depend on the direction the origin is approached in the parameters space $(\nu,d_0,d_2)$.

The action used in this work is restricted to $\ord{\partial_0^2}\ord{\hphi^2}$ and $\ord{\partial_0}\ord{\hphi^4}$ terms and the renormalization of the $\ord{\partial_0^2}$ terms, such as $z$ and $d_2$, were ignored. The resulting Lagrangian with $z=1$ is
\bea\label{ansatz}
L&=&\hf\partial_\mu\phi_+\partial^\mu\phi_+-\hf\partial_\mu\phi_-\partial^\mu\phi_--\frac{m^2}2(\phi_+^2-\phi_-^2)\nn
&&+\nu(\phi_-\partial_0\phi_+-\partial_0\phi_+\phi_-)+id_0(\phi_+-\phi_-)^2+id_2(\partial_0\phi_+-\partial_0\phi_-)^2]\nn
&&-U(\hphi)+V(\hphi)+\hf W_+(\hphi)\partial_0\phi_--\hf W_-(\hphi)\partial_0\phi_+
\eea
where the quartic terms are given by
\bea\label{uvpar}
U(\hphi)&=&\frac{u_{4r}+iu_{4i}}{4!}\phi^4_+-\frac{u_{4r}-iu_{4i}}{4!}\phi^4_-,\nn
V(\hphi)&=&\frac{iv_{2,2}}4\phi^2_+\phi^2_-+\frac{v_{3,1r}+iv_{3,1i}}6\phi^3_+\phi_--\frac{v_{3,1r}-iv_{3,1i}}6\phi_+\phi_-^3.
\eea
Due to the condition $\mr{Re}(S[\hphi])=0$ when $\phi_+=\phi_-$ we use
\be
v_{3,1i}=-\frac{u_{4i}}4-\frac34v_{2,2}
\ee
as a dependent parameter. The wave function renormalization terms to the friction coefficient $W_\pm$ can be simplified by partial integration to
\be\label{wpar}
W_-(\hphi)=\frac{w_{0,3}}6\phi_-^3+\frac{w_{1,2}}2\phi_+\phi_-^2
\ee
with real coefficients and Hermiticity of the density matrix gives $W_+(\phi_+,\phi_-)=W_-^*(\phi_-,\phi_+)$. All parameters appearing in eqs. \eq{uvpar} and \eq{wpar} are real. Yet another inequality to respect is $\mr{Im}(U+V)<0$ to keep the path integral convergent, yielding $u_{22}\ge0$.

\section{Non-relativistic open systems}\label{relboost}
The theory, defined in the previous section violates Lorentz symmetry owing to the time derivatives in the Lagrangian and the spatial momentum cutoff. One wonders whether it is possible to arrive at relativistic open dynamics. We conjecture that only classical open system can be relativistic and quantum open systems always violate relativistic symmetry. The point is that we seek the physical laws governing the observed system which are expressed in terms of differential equations. This latter can be tested only if we have a {\em complete freedom} to choose the initial conditions. Hence the system and its environment must be uncorrelated in the initial state.

\subsection{Classical dynamics}
\begin{figure}
\includegraphics[scale=.5]{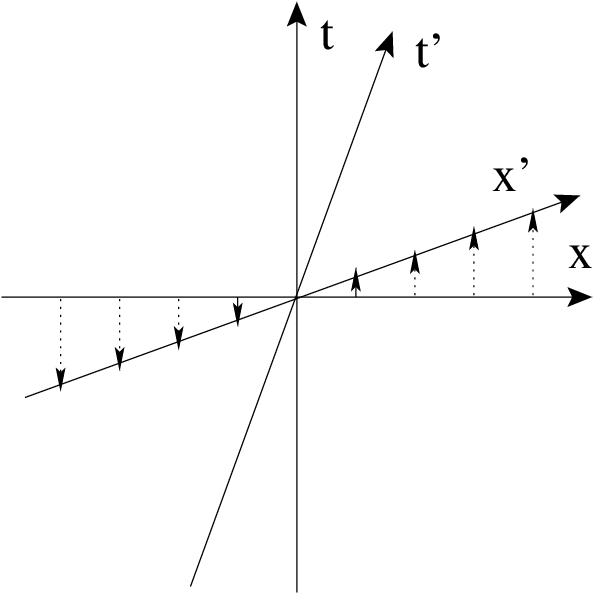}
\caption{Lorentz boost to the right tilts the space-time axes, $x\to x'$, $t\to t'$, and the initial conditions on the $x'$ axis are found by solving the equation of motion along the dotted lines.}\label{boost}
\end{figure}

Let us consider a classical many particle system described by the coordinates $(x_n,y_n,z_n)$, $n=1,\ldots,N$, split it into two parts by the surface $x=0$ and specify the initial conditions on the $t=0$ Cauchy surface. A Lorentz boost in the $x$ direction tilts the $x$ axis and the initial conditions on the new Cauchy surface are found by solving the equations of motion along the dotted lines of Fig. \ref{boost}.

This is a well known complication and it leads to a violation relativistic symmetries and a no-go theorem about the description of relativistic interacting particles in terms of instantaneous interaction potential \cite{currie,leutwyler}. There are ways to arrive at a covariant dynamics by choosing unusual canonical variables but problems arise at the quantization \cite{sazdijan}. The no-go theorem might be considered as the origin of classical fields in physics as the only device to induce relativistic interaction between particles.

Therefore we extend the model by including some classical relativistic field among the dynamical degrees of freedom but continue to consider the surface $x=0$ to separate the two sub-systems of particles. Since interaction between the spatially separated points is excluded by Special Relativity there is no difficulty to keep the initial conditions for the two sub-systems independent during Lorentz boosts.

Perhaps the simplest realistic case is of a single point charge coupled to the electromagnetic field, the latter being considered as environment. One can use independent initial conditions for the charge and the electromagnetic field and the Abraham-Lorentz force is generated as a friction force due to radiation energy loss. The force is fully relativistic and it is proportional to the third derivative with respect to the invariant length because Lorentz invariance excludes Newton's first order form.

\subsection{Quantum dynamics}
In the case of quantum dynamics we assume that the system particles are distinguishable from the environment particles and the initial density matrix at $t=0$ is factorizable in the observed system and its environment. The interaction generated by the time evolution along the dotted lines of Fig. \ref{boost} generates entanglement and the density matrix at $t'=0$ is not factorizable anymore.

The present discussion has some bearing on closed quantum field theory models with fixed, physical UV cutoff. These theories are imagined in a rather vague manner, simply by ignoring completely the degrees of freedom beyond the cutoff. Is this step acceptable in view of the remark that the $i\epsilon$ prescription opens up the dynamics? There is no problem in defining relativistic cutoff theories when smooth cutoff or analytical regularization are employed. But there is no non-perturbative sharp cutoff with Lorentz symmetry. This follows trivially from the observation that a representation of the Lorentz group must include states with arbitrary large momentum. A more transparent argument relies on the non-compact nature of the Lorentz group making the invariant volume of the energy-momentum space between two relativistic hyperbole infinite \cite{boost}.

\subsection{Point charge}
A single charge is a simple but non-trivial case where one can see easily how entanglement arises when relativistic cutoff is used. Let us write the classical effective action as $S=S_0+S_i$ where
\be
S_0=-mc\int ds\left(\sqrt{\dot x^\mu_+\dot x_{\mu+}}-\sqrt{\dot x^\mu_-\dot x_{\mu-}}\right)
\ee
denotes the free action and
\be
S_i=\frac{2\pi e^2}c\int dsds'\hat x^\mu\hat\sigma\hD_{\mu\nu'}(x-x')\hat\sigma\hat x'^{\mu'}
\ee
stands for the particle-field interaction truncated at the quadratic level in the world line \cite{classcharge}. Here $x^\mu=x^\mu(s)$, $x'^\mu=x^\mu(s')$, $x^\mu(s)$ being the world line of the point charge parametrized by the invariant length, $\hD_{\mu\mu'}=-(g_{\mu\mu'}-\partial_\mu\partial_{\mu'})\hD$ is the photon propagator, $\hD$ is a scalar massless propagator \eq{freepro} and $\hat\sigma=\sigma_3$ arises from the structure \eq{herm} of the action. In the quantum case the choice of $S_0$ is different  \cite{feynmanwl,barut,schubert} but $S_i$ remains the same.

To find the physical role of the different blocks of the propagator we write $S_i$ by the help of the parametrization $x_\pm=x\pm x_d/2$ as
\be
S_i=\frac{2\pi e^2}c\int dsds'[\hat x^\mu D^a(x-x')x'_{d\mu}+x^\mu_dD^r(x-x')x'_\mu+ix_d^\mu D^i(x-x')x'_{d\mu}]
\ee
where the block structure \eq{propblock} is used to define the causal propagators and the imaginary part, $D^i$. The first two terms with the causal propagator produce the classical self interaction of the charge and the last, imaginary term describes the decoherence in coordinate basis. Since the initial conditions are identical for $x_+$ and $x_-$ the two classical trajectories are the same $x_d=0$ and $x_d$ plays the simple role of a variational partner of $x$ in completely decohered classical physics \cite{cqco}.

The effective action needs a UV regulator. Since the causal propagators are restricted to the light cone, $D^n(x)=-\delta(x^2)/4\pi$, $D^f(x)=\sign(x^0)D^n(x)$, it is enough to smear the light cone singularity, for instance by the replacement
\be\label{classred}
\delta(z)\to\delta_{\ell_m}(z)=\frac{\Theta(z)}{12\ell_m^4}ze^{-\frac{\sqrt{z}}{\ell_m}}.
\ee
The sign function remains untuched therefore the regulated theory is invariant under $L_\uparrow$, the subgroup of the Lorentz group which leaves the time arrow, inherited from the electromagnetic field, unchanged. What is important is that the symmetry with respect to Lorentz boost is preserved. The near field contribution contains a linear divergence, giving rise of a mass renormalization. The far field contribution is finite but non-uniform convergent as $\ell_m\to0$ and produces the Abraham-Lorentz force in a manner reminiscent of an anomaly in quantum field theories.

However the regularization of $D^i(x)=P1/4\pi^2x^2$, needed in the quantum case, is more cumbersome because it is not localized on the light cone, but relativistic point splitting is available as a perturbative smooth cutoff \cite{classelrad}. The classical regulator \eq{classred} is insufficient in quantum theory owing to decoherence, a signature of system-environment entanglement and parametrized by $D^i$.

\section{Renormalization group for quantum field theories}\label{rgoqft}
The renormalization group method follows the scale dependence of the elementary physical processes of the theory. The scale is defined by the spatial resolution $\ell_m$, the minimal distance of the theory. The blocking step consists of the lowering of the cutoff in small steps, $k\to k-\dk$, and the change of the bare action, $S_k[\hphi]\to S_{k-\dk}[\hphi]$, is obtained by the path integral
\be\label{blocking}
e^{iS_{k-\dk}[\hphi]}=\int D[\hchi]e^{iS_k[\hphi+\hchi]}
\ee
where $\hchi(t,\v{p})$ and $\hphi(t,\v{p})$ are non-vanishing for $k-\dk<|\v{p}|<k$ and $|\v{p}|<k-\dk$, respectively. The energy dimension is chosen in such a manner that the initial value of the gliding cutoff is unity, $k_{in}=1$. The loop expansion to \eq{blocking} gives
\be\label{ev0}
S_k[\phi+\varphi(\phi)]-S_{k-\dk}[\phi]=-\frac{i}2\Tr_{k-\dk<|\v{p}|<k}\ln\fdd{S_k[\phi+\varphi(\phi)]}{\phi}{\phi}
\ee
in the one-loop order where $\varphi(\phi)$ is the saddle point. Since we do not plan to explore condensates the vanishing of the saddle point is assumed. The importance of making the blocking in small steps is that the  $n$-loop correction to this equation is $\ord{(\dk/k)^{n-1}}$ and can be neglected in the limit $\dk\to0$ \cite{wh} where one recovers an exact one-loop evolution equation
\be\label{funceveq}
\frac{d}{dk}S_k[\phi]=-\frac{i}{2\dk}\Tr_{k-\dk<|\v{p}|<k}\ln\fdd{S_k[\phi]}{\phi}{\phi}.
\ee
The right hand side converges because the trace is taken over the function space within a region of volume $\ord{\dk}$ in the momentum space.

\subsection{Substraction point}\label{subtrs}
The handling of functional differential equation is beyond our analytical capabilities hence we project it onto an ansatz functional space, defined by the Lagrangian \eq{ansatz} and convert the resulting evolution equation into a set of coupled differential equations for the paramaters of the Lagrangian. This step contains a choice, the way the parameters are identified in the projected eq. \eq{ev0}. The customary procedure is to define the parameters by evaluating \eq{ev0} at some reference field configuration $\phi_s(x)$. This field will be called subtraction point because this procedure is reminiscent of the subtraction process of the traditional multiplicative renormalization group scheme.

The beta functions $\beta_g=\dot g$ for the parameters of the action are found by identifying the coefficients of certain monomials of the field variable, evaluated at the subtraction point $\hphi(x)=\hphi_0\cos(\omega_sx^0-\v{p}_s\v{x})$. The subtraction point used in previous works was homogeneous, $\v{p}_s=\omega_s=0$, an acceptable choice for imaginary time. Now we use real time and are interested in the physics of realistic, propagating particles thus the subtraction point is placed at the peak of the spectral function, $\omega_s=\omega_p=\sqrt{m^2+\v{p}_s^2}$. One could have chosen the subtraction point somewhere else within the $1/\tau_{diss}$ vicinity of the peak where $\tau_{diss}$ stands for the dissipation time scale, the life-time of the quasi-particles. Our choice, the peak exactly, assumes that the observations in reading off the parameters of the theory last longer time than $\tau_{diss}$ and allows a precise determination of the energies.

To find $\v{p}_s$ recall that the parameters of the action characterize the physics at the cutoff hence the contributions to the beta functions should come from intermediate states $|\v{p}|=k$. The two particles contributing to the beta function of the quartic coupling constant as intermediate states have momentum $\v{p}_1=\v{n}k$, $\v{p}_2=j\v{p}_s-\v{n}k$ with $j\in\{-2,0,2\}$ and we integrate over the unit vector $\v{n}$. The contribution is $\ord{\dk}$ when $j\ne0$ hence we choose the simplest case, $\v{p}_s=0$.

\begin{figure}
\includegraphics[scale=.5]{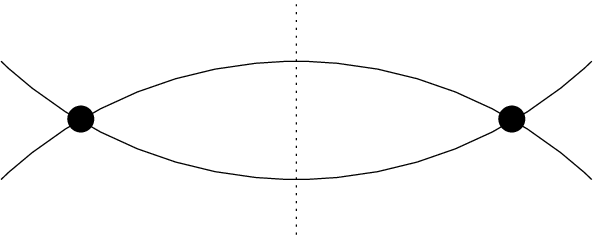}\hskip1cm
\includegraphics[scale=.5]{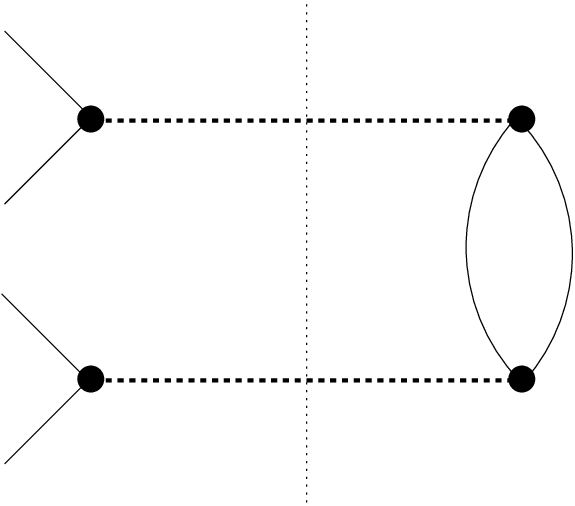}

(a)\hskip4cm(b)
\caption{The Feynman diagramms: (a) $\ord{u_4^*u_4}$ contribution to the beta function of $v_{2,2}$, (b) $\ord{v_{2,2}^2}$ contribution to the beta function of $u_4$. The vertical dotted line separates the bra and the ket factors, the $\phi_+$ and the $\phi_-$ modes, respectively and the horizontal dotted lines connect the $\phi_+^2$ and the $\phi_-^2$ factors in the vertex $v_{2,2}$.}\label{u4inv22}
\end{figure}

Such a subtraction point can pick up the contributions of open processes. In fact, consider the $\ord{u_4^*u_4}$ contribution to the beta function of $v_{2,2}$, depicted in Fig. \ref{u4inv22} (a). The IR particle states with energy momentum $p=(\omega_{k-\dk},\v{0})$ imply the intermediate particle energy momentum $p'_1=(\omega_k,k\v{n})$ and $p'_2=(\omega_k,-k\v{n})$ which generate a system-environment interaction when the MS regulators, the harmonic open parameters, are finite.

\subsection{Beta functions}
The beta functions are found by identifying the coefficient of different multi-linear terms in $\hphi$ on the two sides of the evolution equation \eq{funceveq}. To this end we separate the homogeneous part of $\delta^2S_k[\phi]/\delta\phi\delta\phi$, the free inverse propagator $\hD^{-1}$ given by \eq{fintuneda} and write the evolution equation as
\be
\dot S[\hphi]=-i\frac{k}{2\dk}\Tr\ln\chi[\hD^{-1}-\hat\Sigma]
\ee
where $\dot f=kdf/dk$, the trace is taken over the full functional space and the characteristic function $\chi$ is defined by $\chi(p)=1$ for $k-\dk<|\v{p}|<k$ and $\chi(p)=0$ otherwise. $\hat\Sigma$ denotes the self energy, the second functional derivative of the anharmonic part of the equation arising from the last line of \eq{ansatz}
\be
\Sigma_{\sigma,\sigma'}(p,p')=\sum_{\tau\tau'}\int drdr'\delta(p-p'-r-r')M_{\sigma,\sigma',\tau,\tau'}(p,p',r,r')\phi_\tau(r)\phi_{\tau'}(r')
\ee
with coefficient functions reported in Table \ref{coeff}. The identification of the multi-linear terms in $\hphi$ on both sides of the evolution equation can be carried out after expanding the right hand in the self energy up to second order,
\bea\label{expeeq}
\dot S[\hphi]&=&i\frac{\hbar k}{2\dk}\biggl[\int\frac{dp_1dp_2}{(2\pi)^8}\tr[(\chi\hD)(p_1,p_2)\hat\Sigma(p_2,p_1)]\nn
&&+\hf\int\frac{dp_1dp_2dp_3dp_4}{(2\pi)^{16}}\tr[(\chi\hD)(p_1,p_2)\hat\Sigma(p_2,p_3)(\chi\hD)(p_3,p_4)\hat\Sigma(p_4,p_1)\biggr],
\eea
where a field independent constant has been neglected and $\tr$ denotes the trace is over only the CTP indices.

\begin{table}
\caption{Coefficient function in the self energy, using the notation $v_{3,1}=v_{3,1r}+iv_{3,1i}$.}\label{coeff}
\begin{ruledtabular}
\begin{tabular}{|l|c|r|}
$(\sigma,\sigma')$&$(\tau\tau')=(++)$&$(\tau\tau')=(--)$\\
\hline
\hline
$(+,+)$&$\frac{u_4}2$&$\hf iv_{2,2}$\\
$(-,-)$&$\hf iv_{2,2}$&$-\frac{u_4^*}2$\\
$(+,-)$&$-\frac{i}4w_{0,3}p'^0+\hf v_{3,1}$&$-\frac{i}4w_{0,3}p^0-\hf v_{3,1}^*$\\
$(-+)$&$\frac{i}4w_{0,3}p^0+\hf v_{3,1}$&$\frac{i}4w_{0,3}p'^0-\hf v_{3,1}^*$\\
\end{tabular}
\end{ruledtabular}
\vskip.5cm
\begin{ruledtabular}
\begin{tabular}{|l|c|r|}
$(\sigma,\sigma')$&$(\tau\tau')=(+-)$&$(\tau\tau')=(-+)$\\
\hline
\hline
$(+,+)$&$-\frac{i}4w_{0,3}s^0+\hf v_{3,1}$&$-\frac{i}4w_{0,3}r^0+\hf v_{3,1}$\\
$(-,-)$&$\frac{i}4w_{0,3}r^0-\hf v_{3,1}^*$&$\frac{i}4w_{0,3}s^0-\hf v_{3,1}^*$\\
$(+,-)$&$-\frac{i}4w_{1,2}(p'^0+p^0-r^0+r'^0)+\hf iv_{2,2}$&$-\frac{i}4w_{1,2}(p'^0+p^0-r'^0+r^0)+\hf iv_{2,2}$\\
$(-,+)$&$\frac{i}4w_{1,2}(p^0+p'^0+r^0-r'^0)+\hf iv_{2,2}$&$\frac{i}4w_{1,2}(p^0+p'^0+r'^0-r^0)+\hf iv_{2,2}$\\
\end{tabular}
\end{ruledtabular}
\end{table}

The beta function of a parameter $g$ is defined by the help of a differential operator $B_g$, given in terms of the functional derivative with respect to $\hphi(\omega,\v{p})$,
\bea
B_{m^2-id_0}&=&\frac1{\delta^{(4)}(0)}\fdd{}{\phi_+(-\omega_s,\v{0})}{\phi_+(\omega_s,\v{0})},\nn
B_\nu&=&i\partial_{\omega_s}\frac1{\delta^{(4)}(0)}\fdd{}{\phi_-(-\omega_s,\v{0})}{\phi_+(\omega_s,\v{0})},\nn
B_{u_4}&=&-\frac1{\delta^{(4)}(0)}\frac{\delta^4}{\delta\phi^2_+(-\omega_s,\v{0})\delta\phi^2_+(\omega_s,\v{0})},\nn
B_{v_{2,2}}&=&-\frac1{\delta^{(4)}(0)}\frac{\delta^4}{\delta\phi^2_-(-\omega_s,\v{0})\delta\phi^2_+(\omega_s,\v{0})},\nn
B_{v_{3,1}}&=&-\frac1{\delta^{(4)}(0)}\frac{\delta^4}{\delta\phi_-(-\omega_s,\v{0})\delta\phi_+(-\omega_s,\v{0})\delta\phi_+^2(\omega_s,\v{0})},\nn
B_{w_{1,2}}&=&i\partial_{\omega_s}\frac1{\delta^{(4)}(0)}\frac{\delta^4}{\delta\phi^2_-(-\omega_s,\v{0})\delta\phi_+(\omega_s,\v{0})\delta\phi_+(-\omega_s,\v{0})},\nn
B_{w_{0,3}}&=&i\partial_{\omega_s}\frac{\delta^4}{\delta\phi^2_-(-\omega_s,\v{0})\delta\phi_+(\omega_s,\v{0})\delta\phi_+(-\omega_s,\v{0})},
\eea
where $\delta^{(4)}(0)=(2\pi)^4/V^{(4)}$ is the value of the regulated Dirac delta in energy-momentum space belonging to the quantization space-time volume $V^{(4)}$. We apply $B_g$ on both sides of the evolution equation \eq{expeeq} and make the replacement $\hphi(x)=\hphi_s(x)$ in the resulting equation.

The beta functions $\beta_g$ obtained in such a manner correspond to the momentum-independent constant in the free action \eq{fintuneda}. Note that $z$ and $d_0$ do note evolve. The beta function for $\tilde\nu$ is
\be
\beta_{\tilde\nu}=\hf\omega_k\beta_\nu+\frac{k\tilde\nu}{2\omega_k}.
\ee

\subsection{Complex parameters}
Note that the action used in the path integral formalism is not the classical one. It has been mentioned in section \ref{conds} that real actions lead to Hermitian Hamiltonian in closed dynamics. But the reality of the action is only a sufficient rather than a necessary condition, the latter being unknown. The use of a subtraction point within the mass-shell of a closed theory makes the renormalized parameters, given in terms of one-particle irreducible diagrams at the subtraction point, complex owing to the optical theorem. This opens the way that the action possess a finite imaginary parts.

The finite imaginary part, a possibility for closed interaction, is a necessity for open dynamics in a soft environment: The dynamical breaking of the time reversal symmetry by dissipation generates finite life-time for the quasi-particles. The action for open dynamics is richer than its closed counterpart not only due to the reduplication of the dynamical variables but owing to the unavoidable imaginary part of certain parameters. In fact, the imaginary part of a parameter has the same superficial degree of divergence than the real part therefore there are more renormalizable parameters. It is worthwhile noting that the optical theorem, indicating the complex nature of the parameters of the action, is already about open dynamics.

\section{Phase transitions, scaling laws}\label{scalaw}
The phase structure and the characteristic scale of the theory \eq{ansatz}-\eq{uvpar} are discussed here by inspecting the renormalization group flow. Instead of covering the complete space of parameters we focus on weakly open theories, where the parameters at the initial cutoff, called bare parameters, are close to an open theory. The quadratic bare open parameters were small but non-vanishing to regulate the MS divergences and $u_{4iB}$ together with the open coupling constants $v_{2,2B}$, $v_{3,1B}=v_{3,1rB}+iv_{3,1iB}$, $w_{0,3B}$ and $w_{1,2B}$ were vanishing for the sake of simplicity.  All open parameters of the theory reach finite strength during the evolution, the dynamics becomes open when the cutoff is moved. The opening of the theory starts with the contribution of the Feynman graph of Fig. \ref{u4inv22} (a) generating finite $v_{2,2}$ which in turn ignites the evolution of the remaining open parameters.

Two phase transitions can be seen in the numerical integration of the evolution equations: One is the usual second order transition related to the spontaneous breakdown of the $\phi\to-\phi$ $Z_2$ symmetry, not explored in this work. Another transition, studied in more detail separates phases with weakly and strongly open dynamics, in particular with regular and strong decoherence, respectively. Three crossovers are identified, the relativistic crossover, the opening crossover and the correlation length.

\subsection{Relativistic crossover, non-relativistic scaling}
Massive theories in Euclidean space-time display a crossover separating the UV and the IR scaling laws. It is found where the minimal distance reaches the correlation length, $\ell_m=\xi_E=1/m$. The IR scaling is trivial, the only relevant parameter being the mass. In fact, the field variable can be interpreted as the average
\be
\phi(t,\v{x})=\frac3{4\pi\ell_m^3}\int d^3y\Theta(\ell_m-|\v{y}|)\phi(t,\v{x}+\v{y})
\ee
over a sphere of radius $\ell_m$. Close to the IR end point, $k\to0$, $\ell_m$ is large and the inequality $\ell_m\gg\xi_E$ makes the blocked field dominated by statistically independent contributions, the condition to apply the central limit theorem stating that the probability density is Gaussian.

\begin{figure}
\includegraphics[scale=.5]{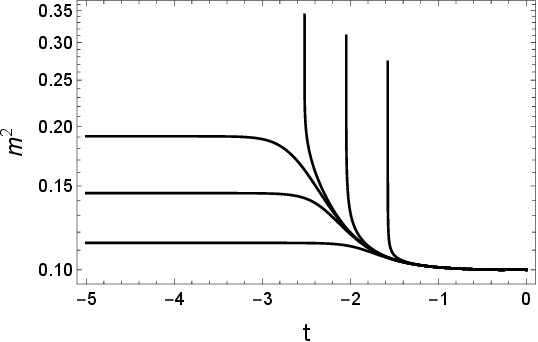}\hskip1cm
\includegraphics[scale=.5]{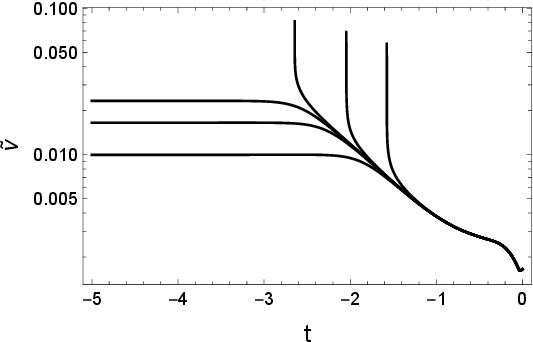}\hskip1cm
\includegraphics[scale=.5]{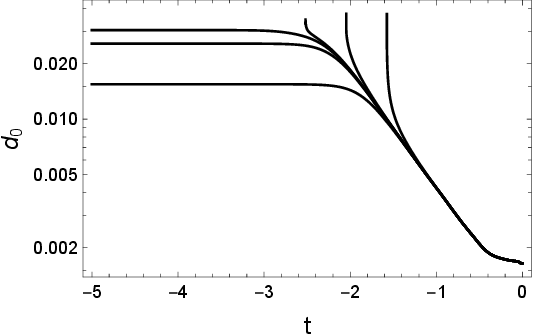}

\includegraphics[scale=.5]{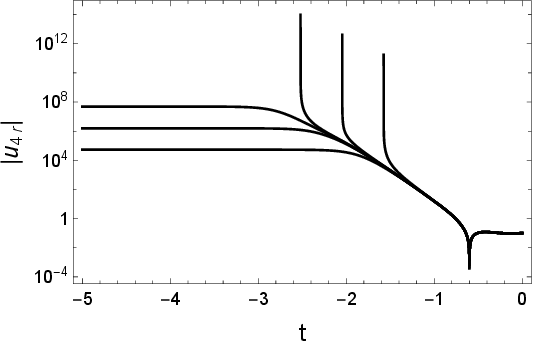}\hskip1cm
\includegraphics[scale=.5]{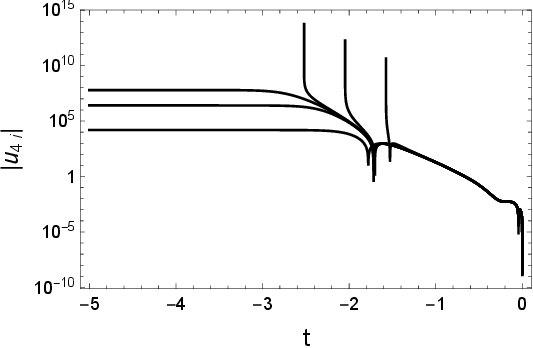}\hskip1cm
\includegraphics[scale=.5]{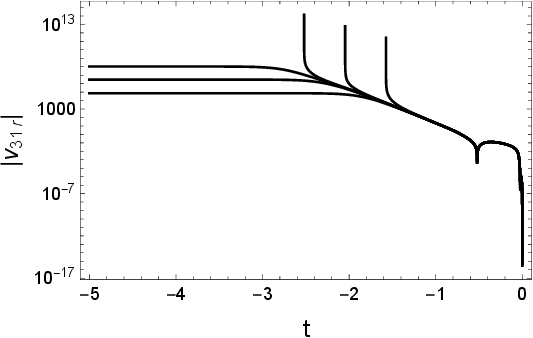}

\includegraphics[scale=.5]{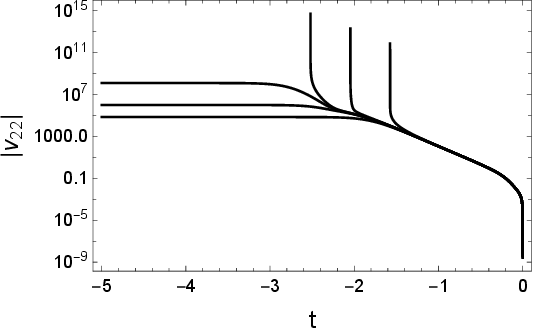}\hskip1cm
\includegraphics[scale=.5]{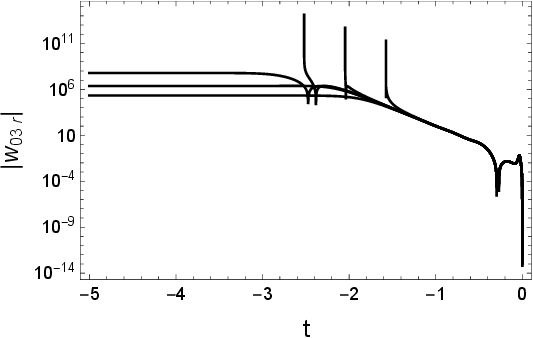}\hskip1cm
\includegraphics[scale=.5]{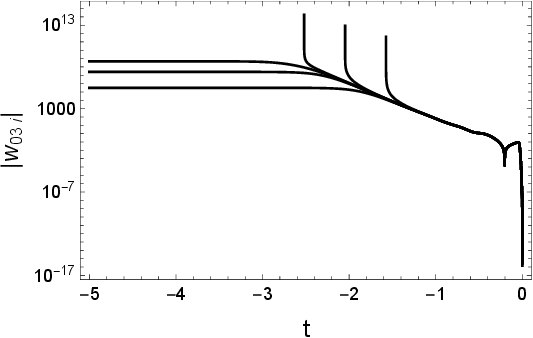}

\includegraphics[scale=.5]{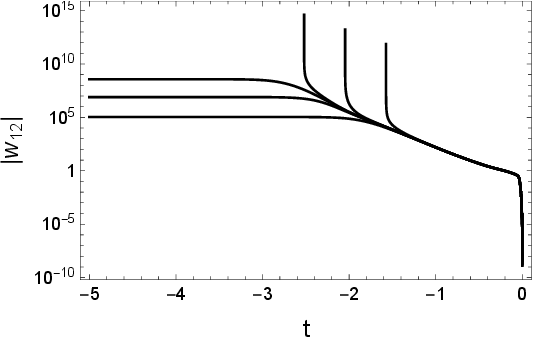}
\caption{Renormalized trajectories close to the decoherence phase transition. The initial conditions are $m^2_B=0.1$, $\tilde\nu_B=0.1g^2$, $d_{0B}=0.1g^2$, $\tilde d_{2B}=0.1g^2$, $u_{4rB}=0.1$. Different trajectories belong to the values $g^2\in[0.016478,0.0164788]$.}\label{sepf}
\end{figure}

The UV scaling regime ends at a crossover in Minkowski space-time theories, as well, but this is a relativistic crossover because $k_{rel}=m(k_{rel})$ indicates the onset of non-relativistic physics. The observation of the slow non-relativistic motion needs $x^0\gg|\v{x}|$ which in turn introduces a better energy resolution, $E=m+E_{nr}$, $E_{nr}\ll E$, and opens up the well known richness of non-relativistic phenomena. The relativistic crossover is about $k_{rel}\approx-0.5$ in Fig. \ref{sepf} and a non-relativistic scaling regime can clearly be identified in the scale interval $-2.5<t<-0.5$. The highly non-trivial nature of the non-relativistic physics is reflected in the strong evolution of the coupling strengths. The quadratic and quartic parameters follow qualitatively different trajectories. While the harmonic open parameters remain weak the open coupling constants, anharmonic parameters, follow a faster, power law evolution.

The non-relativistic regime can be weakly or strongly coupled. Let us first restrict our attention to bare theories with infinitesimal open couplings. When the crossover is reached by a sufficiently weak $u_{4r}$ then the theory remains weakly coupled along the continuation of the renormalization group flow. However if $u_{4r}$ is larger than a critical value, depending on the other parameters, then the non-relativistic scaling laws generate strong couplings when two conditions met.

To find these conditions we start with the non-relativistic beta function of $u_{4r}$ in the STP formalism by ignoring the imaginary parts, $\dot u_{4r}=\beta_{u_{4r}}k^3u_{4r}^2$ with $\beta_{u_{4r}}=9/2^7\pi^2m^3$. Its solution,
\be\label{stpevol}
u_{4r}=\frac{u_{4r0}}{1-\frac{\beta_{u_{4r}}}3u_{4r0}(k^3-k_0^3)},
\ee
decreases $u_{4r}$ but keeps it positive. However there are other CTP contributions in the beta function, in particular an $\ord{v_{2,2}^2}$ term, shown on Fig. \ref{u4inv22} (b). This contribution is positive and drives $u_{4r}$ across zero if $v_{2,2}$, generated so far by $u_4$, is sufficiently strong. This is the first condition. When $u_{4r}$ crosses zero with sufficiently weak open couplings the theory is weakly coupled and retains this feature in the ensuing evolution. The second condition is to have sufficiently strong $v_{2,2}$ at the flipping of the sign of $u_{4r}$ which can drive $u_{4r}$ to strongly negative values with the power law scaling of eq. \eq{stpevol}.

\begin{figure}
\includegraphics[scale=.5]{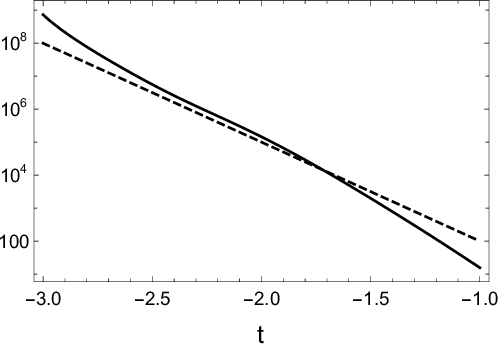}
\caption{Running of $u_{4r}$ in the non-relativistic regime (solid line) and the STP evolution \eq{stpevol} (dashed line).}\label{nonrelf}
\end{figure}

\begin{figure}
\includegraphics[scale=.6]{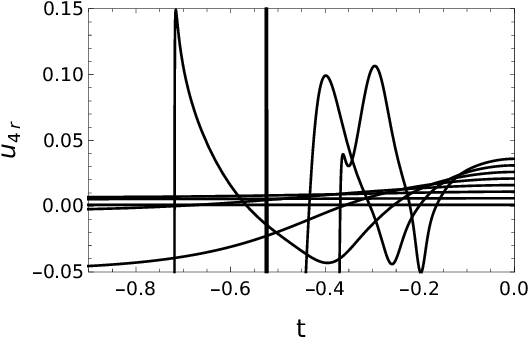}\hskip1cm
\includegraphics[scale=.6]{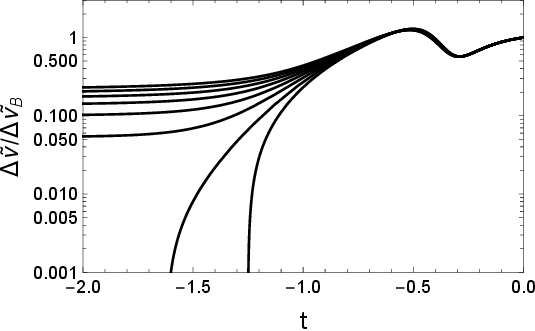}

(a)\hskip4cm(b)
\caption{The crossing of the trajectories. (a): The trajectory of $u_{4r}$ for $m^2_B=0.1$, $g=1$, $\tilde\nu_B=d_{0B}=\tilde d_{2B}=0.01$, $u_{4rB}\in[10^{-3},7\times10^{-3}]$. The three largest $u_{4r}$ trajectories are in the strongly open phase.(b): The finite difference $\Delta\tilde\nu/\Delta\tilde\nu_B$ for $m^2_B=10^{-4}$, $d_{0B}=\tilde d_{2B}=0.1$, $g^2\in[0.166,0.167]$. The trajectories are in the weakly open phase, the finite difference is increasing with $g$ and changes sign at the lowest two $g$ values.}\label{crossf}
\end{figure}

The competition of $\ord{u_4^*u_4}$ and $\ord{v_{2,2}^2}$ terms in $\beta_{u_4}$ may lead to the crossing of the trajectories of $u_{4r}$, shown on Fig. \ref{crossf} (a). In fact, stronger $u_{4r}$ generates stronger $v_{2,2}$ which in turn pushes $u_{4r}$ faster to negative values. Such an exchange of the strongly and the weakly coupled dynamics takes place around $k_{rel}$.

To close the discussion of the relativistic crossover we note that there is another, simpler mechanism in this crossover regime which can map strongly (weakly) open UV dynamics onto weakly (strongly) open IR dynamics. This is due to a multiplicative factor $1/\nu$ in the beta functions produced by the application of the residuum theorem to the energy integral. The theories with smaller $\nu$ tend to have large beta functions which drives the open parameters to larger values. The crossing of the $\tilde\nu$ trajectories is captured numerically on Fig. \ref{crossf} (b) by considering a family of trajectories $\tilde\nu_j$ with initial conditions $\tilde\nu_{Bj+1}=\tilde\nu_{Bj}+\Delta\tilde\nu_B$. The neighboring trajectories cross when the sign of $\partial\tilde\nu_j/\partial\tilde\nu_B\approx(\tilde\nu_{j+1}-\tilde\nu_j)/\Delta\tilde\nu_B$ flips.

\subsection{Open-closed phase transition}\label{openptrs}
When both $u_{4r}$ and $v_{2,2}$ are sufficiently strong to reach a strongly coupled non-relativistic scaling then a further, third condition shapes the IR dynamics. Let us first look at the numerical evidences. The initial conditions of the renormalized trajectories of Fig. \ref{sepf} approach the closed initial theory along a straight line in the open quadratic parameter space, $(\tilde\nu_B,d_{0B},\tilde d_{2B})=(g^2\bar\nu_B,g^2\bar d_{0B},g^2\bar d_{2B})$, parametrized by the system-environment coupling strength $g^2$. We find a separatrix reminiscent of the scaling around the Wilson-Fischer fixed point in an Euclidean theory under four dimensions. The fine tuning of $g^2$ allows us to push the separation of the trajectories arbitrary close to $k=0$. The emerging long non-relativistic scaling regime in $t=\ln k/k_{in}$ with qualitatively different scaling laws on the two sides of the separatrix induces non-analitical dependence of the IR observables on the UV parameters, the condition of a phase transition. Similar phase transition is found by approaching the origin along others straight lines. Therefore the theory supports an environment induced phase transition where the strength of open interaction channels changes in a singular manner. The trajectories of Fig. \ref{sepf} belong to momentum-dependent parameters $\tilde\nu$ and $\tilde d_2$ but the renormalization group flow obtained by using momentum-independent parameters $\nu$ and $d_2$ is very similar.

Look first into the weak $g^2$ side of the separatrix where the trajectories start to diverge at a scale $k=k_o$ where the open interaction channels become suddenly strong. Only a short evolution is reliable beyond this crossover scale, slightly below $t=-1.5$, since our ansatz \eq{ansatz}, build on the Landau-Ginzburg double expansion of low order, is not justifiable for strong coupling theories. The lowering of the cutoff always produces a well defined blocked theory therefore the divergence of the trajectories, an IR Landau pole around $k_o$, indicates the insufficiency of our ansatz. The allowed parameters have to diverge to assure the renormalization group invariant physics. It was checked numerically that the renormalized trajectory remains finite if $w_{0,3}$ is frozen during the evolution. Hence the singularity driven by the open interaction channels.

Our ansatz is sufficient to describe the phase at the strong $g^2$ side of the separatrix since the renormalized trajectories remain finite down to the IR end point $k=0$. Therefore the open interaction channels are sufficient and the IR physics should be more closed in this phase.

The trajectory crossing, needed to exchange the weakly and the strongly  open theories during the evolution, is observed at arbitrary $u_{4rB}$, smaller $u_{4rB}$ leads to smaller $g^2$ at the separatrix. The crossing extends over small enough $g^2$, the largest value being slightly within the weakly open phase. The decrease of $m^2_B$ and the increase of the relativistic scaling regime process result in a more involved scaling laws around the separatrix and makes $u_{4r}$ changing sign several times but the trajectory crossing remain located around $k_{rel}$.

The possibility of creating a long evolution close to the separatrix by the fine tuning of the initial conditions is a characteristic feature of second order phase transitions and results from competing effects. In the case of the spontaneous breakdown of the $\phi\to-\phi$ symmetry $m^2<0$ indicates an attraction between the particles in the empty, symmetric vacuum which competes with the repulsive force arising from collisions, represented by the $\phi^4$ vertex. The beta functions of the quartic couplings are sums of terms, each being the product of two coupling constants and a loop integral. The long non-relativistic scaling regime along the separatrix of the open-closed phase transition results from the competition of the positive and the negative terms.

\subsection{Correlation length}\label{centrlth}
\begin{figure}
\includegraphics[scale=.5]{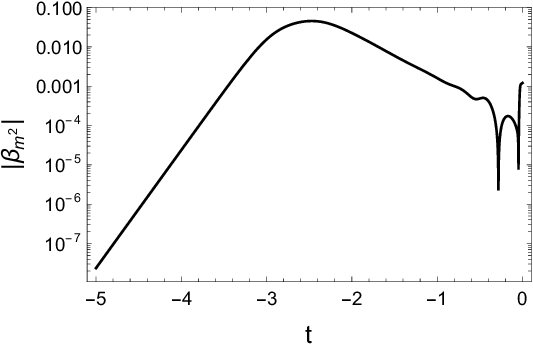}\hskip1cm
\includegraphics[scale=.5]{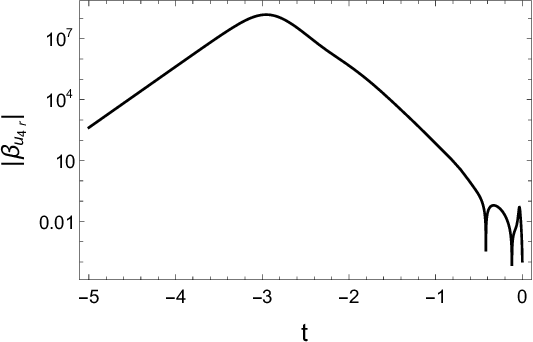}\hskip1cm
\includegraphics[scale=.5]{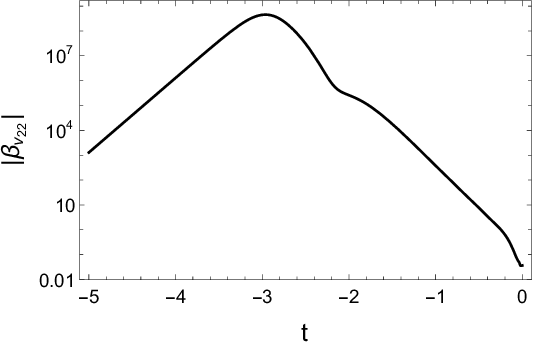}

(a)\hskip5cm(b)\hskip5cm(c)
\caption{The beta functions of (a) $m^2$, (b): $u_{4r}$ and (c): $v_{22}$ along the trajectory of Fig. \ref{sepf} belonging to the fifth value of $g^2$ with the longest non-relativistic scaling and $k$-independent trajectories for $t<t_{corr}\approx-3$. The relativistic crossover is at $t_{rel}=-0.5$.
}\label{nrscbf}
\end{figure}

Finally we turn to the strong $g^2$ side of the separatrix where the trajectories remain regular down to $k=0$. One finds a new crossover scale $k_{corr}$ in Fig. \ref{sepf} where the evolution seems to stop. However this is a too hasty conclusion because the beta functions depend on $k$, too, and we need $\beta=0$ and  $\dot\beta=0$, an overdetermined system of equations, for the fixed point. Actually the evolution only slows down and we witness the build up of an intricate relation between the parameters, namely, the beta functions become proportional to $k^3$, c.f. Fig. \ref{nrscbf} for few selected examples.

The importance of the $\ord{k^3}$ scaling of the beta functions is that it makes the right hand side of eq. \eq{funceveq} $\ord{k^2}$. The functional trace is calculated in the momentum space where it appears as an integral over a thin spherical shell of radius $k$ and thickness $\dk$. The integrand is a finite constant owing to the rotational symmetry of the subtraction point. The factor $k^2$ in question is just the integration measure thus the integrand is {\em $k$-independent}. A good approximation of the inverse propagator, $\hD^{-1}(\omega,\v{p})$, at the scale $|\v{p}|=k$ is provided by the second functional derivative of the action, given by eqs. \eq{diblock}-\eq{fintuneda} with parameters read off from the renormalized trajectories, evaluated  at $\phi_\pm=\la\phi\ra=0$. The correlation length is usually defined as a length scale $\xi=1/k_{corr}$ beyond which $\hD^{-1}$ is momentum-independent. We found the inequality $k_{corr}<k_{rel}$ to hold in the numerical examples suggesting that the correlation length is always non-relativistic.

The interpretation $\xi=1/k_{corr}$ can be corroborated by the considering $\phi(x)$ as a random variable averaged over a block of volume $\ell_m^3$. Such a spatial blocks are not sharply defined, $\phi(x)$ is actually defined as a local variable in the continuous three-space. What is important is that its Fourier transform is vanishing for $|\v{p}|>k$ thus the values of $\phi(x)$ at distance less than $\ell_m$ can not differ strongly, $\ell_m<\xi$. How can we notice during the blocking that we reached the correlation length with our increasing minimal length, $\ell_m=\xi$? Since the generally expected relation $\xi=1/m$ can not be applied to non-relativistic theories we use the central limit theorem to read off $\xi$ from the renormalized trajectories. The theorem states that the average of $N$ independent random variables $\phi_n$, $n=1,\ldots,N$, with identical probability distribution and finite second moment $\sigma$ follows a Gaussian distribution function located at $\la\phi\ra$ with spread $\sigma/\sqrt{N}$ and can be generalized to quantum systems \cite{macr}. The vision of $\phi(x)$ at a certain cutoff $k$ as a random variable smeared over a volume $\ell^3_m$ suggests that the lowering the cutoff $k\to k/s$ leads to field variables averaged over a volume $(s\ell)^3_m$. Thus the correlation length can be defined by the requirement that the second moment of the field variable be proportional to $1/\ell_m^{3/2}$ for $k<1/\xi$.

The square of the second moment of the field variable can be expressed in terms of the propagator,
\be\label{phisecmom}
\la\phi^2(x)\ra=\int\frac{d\omega d^3p}{(2\pi)^4}\Theta(k-|\v{p}|)iD_{\sigma\sigma'}(\omega,\v{p})
\ee
with $\hD^{-1}\approx\delta^2S_k[\hphi]/\delta\hphi\delta\hphi_{|\hphi=0}$. The momentum integral is easy to calculate by the help of the approximation $\omega_p\approx m$, valid for $k<k_{rel}$,
\be\label{secmomp}
\la\phi^2(x)\ra=-D^i(0)=\frac{n_0}{12\pi^2m\ell^3_m}
\ee
where $n_0=(d_0+d_2m^2)/\nu m$ is a dimensionless constant. The proportionality of the second moment with  $1/\ell^3_m$ is consistent with the interpretation of $\xi=1/k_{corr}$ as correlation length where the quantum fluctuations become scale-independent.

\subsection{Critical behaviors}
The $\phi^4$ model in Euclidean space-time supports critical behavior in the vicinity of the massless case where the renormalized mass, the parameter $m^2$ at the IR end-point $k=0$, is vanishing and the singularities and critical exponents are generated by the long UV scaling regime. Such a simple picture is valid only if the IR scaling regime trivial and can not produce long non-trivial scaling regime.

\begin{figure}
\includegraphics[scale=.5]{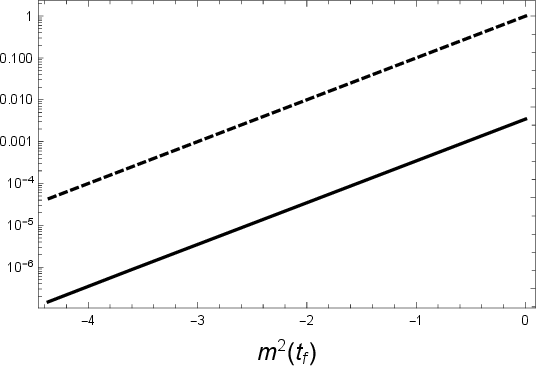}
\caption{The crossover scales $k^2_{rel}$ (upper) and $k^2_{corr}$ (lower) as the function of the renormalized mass, read off at the IR end $t_f=-5$ of the renormalized trajectories where it does not evolve anymore, at $\tilde\nu_B=d_{0B}=\tilde d_{2B}=u_{4rB}=0.1$ with different choice of $m_B^2$.}\label{scales}
\end{figure}

To check the traditional critical behavior we performed the limit $k_{rel}\to0$. A practical definition of the crossover scale $k_{corr}$ used in the calculation is the scale $k$ where $\dot\beta/\beta$ reaches $3$ with a precision of 1\%. The resulting numerical value varies slightly on the choice of the beta function. The crossover scales $k^2_{corr}$ and $k^2_{rel}$ are shown on Fig. \ref{scales} as the function of the renormalized mass $m^2(k)$ taken at $k\sim0$. The curves support a linear relation, $k_{corr}=c_{corr}m_R$, $k_{rel}=c_{rel}m_R$ with $c_{corr}<c_{rel}$ and the usual scenario of critical phenomena is recovered in the static sector of the theory in Minkowski space-time.

The theory displays another critical behavior, as well. Close to the separatrix of Fig. \ref{sepf} $k_{corr},k_o\to0$ and $\xi\to\infty$. Note that the critical behavior around the separatrix with finite, non-vanishing $k_{rel}$ belongs to an IR rather then UV fixed point.

\section{Summary}
One of the basic problems of quantum field theory is the understanding of the transition to classical field theories at macroscopic spatial resolution. Such a macroscopic limit is outlined here for a simple model of a self interacting scalar field. The spatial resolution is controlled by the sharp UV cutoff in momentum space. The macroscopic limit is driven by the short distance modes beyond the UV cutoff which play the role of an environment and generate an open dynamics for the resolved modes. It is argued that the UV cutoff which separates the observed system and its environment must be non-relativistic.

The physics of the theory is inferred from its renormalization group flow. To identify the dynamics of propagating quasi-particles one has to chose a subtraction point where the running parameters are read off close to the peak of the spectral function. The main difficulty of such a subtraction method is the $1/\nu$ multiplicative factor, a MS divergence, of the beta functions. Hence one has to start small but non-vanishing open quadratic parameters and the closed limit can be realized only numerically. This limit all the more difficult beacuse the evolution equation becomes stiff for small $\nu$.

The UV scaling regime ends at the Compton wavelength beyond which the physics is non-relativistic. The non-relativistic scaling laws can generate strong couplings in the IR by pressing the closed self interaction strength to negative values and a second order phase transition is found separating strongly and weakly open theories. Note that the open interaction always implies decoherence. The crossing of the trajectories around this phase transition realizes a counter-intuitive dynamics where the strongly (weakly) open UV theories evolve into a weakly (strongly) open IR dynamics. We can not see the IR limit of the strongly open phase owing to our limited ansatz for the blocked action. There is no such problem in the weakly open phase where a further crossover is found indicating that a non-relativistic correlation length is generated which is longer the Compton wavelength.

Our main result is that a closed bare theory is strongly decohered in the IR and its quantum fluctuations beyond the correlation length are suppressed by the central limit theorem. It is remarkable that all this is achieved in a theory with a gap in its excitation spectrum by the help of the unresolved UV modes without any external environment. However several questions are left for further studies. (i) It would be important to find the analytical form of the blocked action which allows us to reconstruct the IR dynamics within the strongly open phase.  (ii) Another second order phase transition is found beyond the usual order-disorder transition. Is there a symmetry and an order parameter to characterize this transition? What is the universality class of this phase transition? (iii) Only weakly open bare theories are considered in this work. Are there further phase transitions for genuinely open quantum field theories? (iv) The phase transition between weakly and strongly open theories makes the comparison of quantum and classical models more involved. In fact, ever since the discovery of the quark confinement problem in QCD one is accustomed to situations where the dynamical degrees of freedom change with the resolution. Do we have qualitatively similar quasi-particles in the quantum and the classical domain of quantum field theories?

\end{document}